\documentclass[apjl,numberedappendix,letterpaper]{emulateapj}
\usepackage[T1]{fontenc}
\usepackage{graphicx}
\usepackage{ulem}

\makeatletter

\usepackage[T1]{fontenc}
\slugcomment{}
\shorttitle{}
\shortauthors{M. Wolleben et al.}

\def\radm2{rad~m$^{-2}$}

\newcommand{\Brpa}{B_{\parallel}}
\newcommand{\Brpe}{B_{\perp}}

\newcommand{\n}{n_\mathrm{e}}
\newcommand{\RM}{\mathrm{RM}}

\newcommand{\cmcube}{\,\mathrm{cm^{-3}}}

\renewcommand{\deg}{^{\circ}}
\newcommand{\sqdeg}{\,\mathrm{deg}^2}

\newcommand{\FRM}{\,\mathrm{rad\,m^{-2}}} 

\newcommand{\kms}{\,\mathrm{km\,s^{-1}}}

\newcommand{\pc}{\,\mathrm{pc}}

\newcommand{\uG}{\,\mu\mathrm{G}} 

\begin{document}

\title{Antisymmetry in the Faraday Rotation Sky Caused by a Nearby Magnetized Bubble}

\author{M. Wolleben\altaffilmark{1,12}}
\author{A. Fletcher\altaffilmark{2}}
\author{T. L. Landecker\altaffilmark{1}}
\author{E. Carretti\altaffilmark{4}}
\author{J. M. Dickey\altaffilmark{5}}
\author{B. M. Gaensler\altaffilmark{6}}
\author{M. Haverkorn\altaffilmark{3,11}}
\author{N. McClure-Griffiths\altaffilmark{4}}
\author{W. Reich\altaffilmark{9}}
\author{A. R. Taylor\altaffilmark{10}}
\email{maik.wolleben@nrc.gc.ca}

\altaffiltext{1}{Dominion Radio Astrophysical Observatory, HIA-NRC, Canada}
\altaffiltext{2}{School of Mathematics and Statistics, Newcastle University}
\altaffiltext{3}{ASTRON}
\altaffiltext{4}{ATNF - CSIRO Astronomy and Space Science, Australia}
\altaffiltext{5}{Physics Department, University of Tasmania, Australia}
\altaffiltext{6}{Sydney Institute for Astronomy, School of Physics, The University of Sydney, NSW 2006, Australia}
\altaffiltext{9}{Max-Planck-Institut f\"ur Radioastronomie, Germany}
\altaffiltext{10}{Centre for Radio Astronomy, University of Calgary, Canada}
\altaffiltext{11}{Leiden Observatory, PO Box 9513, 2300 RA Leiden}
\altaffiltext{12}{Covington Fellow} 

\begin{abstract}
Rotation measures of pulsars and extragalactic point sources have
been known to reveal large-scale antisymmetries in the Faraday rotation
sky with respect to the Galactic plane and halo that have
been interpreted as signatures of the mean magnetic field in the
Galactic halo. We describe Faraday rotation measurements of the diffuse Galactic polarized radio emission over a large region in the northern Galactic hemisphere. Through application of Rotation Measure Synthesis we achieve sensitive Faraday rotation maps with high angular resolution, capable of revealing fine-scale structures of $\sim 1$\degr~in the Faraday rotation sky. Our analysis suggests that the observed antisymmetry in the Faraday rotation sky at $b>0^\circ$ is dominated by the magnetic field around a local HI bubble at a distance of ~100 pc, and not by the magnetic field of the Galactic halo. We derive physical properties of the magnetic field of this shell, which we find to be 20 - 34$\uG$ strong. It is clear that the diffuse polarized radio emission contains important information about the local magneto-ionic medium, which cannot yet be derived from Faraday rotation measures of extragalactic sources or pulsars alone.

\end{abstract}


\section{Introduction}

Linearly polarized radio emission is Faraday rotated as it passes through the interstellar medium (ISM) by an angle
\begin{equation}
\Delta\theta=\phi\lambda^2=0.81\int^{near side}_{far side}\n(l)\Brpa(l)\mathrm{d}l \ \lambda^2,
\label{eq:FD}
\end{equation}
where $\phi$, called the Faraday depth, is the integral along the line-of-sight (l.o.s.) $l/[\pc]$ of the product of the free electron density $n_e/[\cmcube]$ and the component of the magnetic field along the l.o.s., $\Brpa/[\uG]$. For wavelengths ($\lambda/[\mbox{m}]$) greater than a few cm, $\Delta\theta$ is a few degrees or more for typical interstellar values of $\n\approx 0.1\cmcube$ and $\Brpa\approx 1\uG$ along kpc pathlengths. Thus, in principle, the Faraday rotation of polarized emission can be used to identify the structure of the magnetic field of the Milky Way. 

The Faraday rotation of (mainly unresolved) polarized extragalactic sources (EGS) and pulsars can be readily measured but provides an irregularly sampled, sparse grid of $\phi$ which may include contributions from Faraday rotation within the sources. In the case of EGS the Faraday rotation arising in the Galactic magneto-ionic medium is integrated along the entire l.o.s. through the Milky Way, while using pulsars one can try to compare the Faraday rotation from pulsars that are nearby on the sky but lie at different distances to sample the magnetic field in localized regions. In either case, the existence of small-scale, high-RM structures in the Galactic neighbourhood makes it difficult to distinguish the contributions of large local features, such as the North-Polar Spur (NPS), from the global Milky Way magnetic field structure. The interpretation of the data requires careful statistical analysis and modelling to try to tease apart these effects.

In the case of polarized synchrotron emission originating from the diffuse ISM, the Faraday rotation is uniformly sampled at the scale of the telescope beam but a major problem is that Faraday rotation can depolarize the emission, with a non-linear dependence on frequency, so $\phi$ cannot always be reliably identified with the rotation measure 
\begin{equation}
\RM=\Delta\theta/\Delta\lambda^2,
\label{eq:RM}
\end{equation}
obtained from observations at different wavelengths \citep[see][]{1966MNRAS.133...67B, Sokoloff:1998}. This can be partly overcome using Rotation Measure Synthesis \citep[RM-synthesis, ][]{2005A&A...441.1217B} with observations that contain many frequency channels across a wide band. Choosing a value for $\phi$, the observed polarization vector in each frequency channel is derotated by ${\phi}{\lambda^2}$ and the rotated vectors are coherently added to obtain an image of polarized intensity at that value of Faraday depth. The final product of RM-synthesis is a 3-D data cube (of polarized emission) with the first and second dimensions being the coordinates on the sky, and the third dimension being Faraday depth, $\phi$ [\radm2]. Faraday depth is usually not related to physical distance.

We have produced an RM-synthesis cube of a large field toward the Galactic centre, using new wide-band data obtained with the 26-m Telescope of the Dominion Radio Astrophysical Observatory (DRAO). In this paper we present evidence that a nearby HI bubble produces a large-scale RM pattern on the sky that mimics the effect that an overall antisymmetry in the Milky Way's regular magnetic field would have on EGS and pulsar Faraday rotation, both in magnitude and sign. Our results demonstrate the power of RM-synthesis and high resolution polarization surveys in separating global from local magnetic field structures in the Milky Way.

\section{Observations}

\begin{figure}[tbp]
  \includegraphics[clip, width=\columnwidth, angle=0, bb=52 309 638 713]{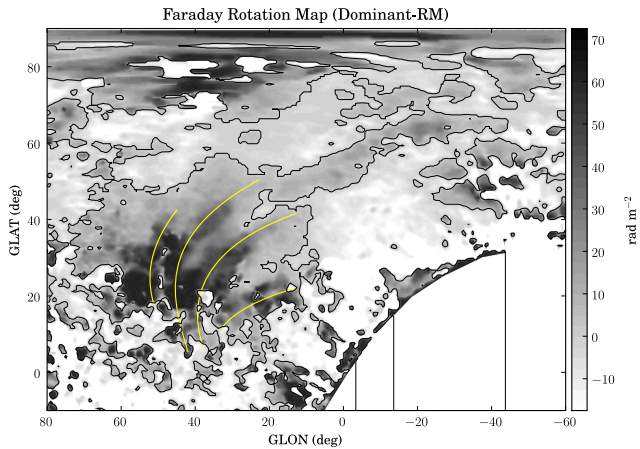}
  \caption{A Faraday rotation map showing the Faraday depth $\phi$ of the dominant emission in each pixel (the position of the peak in the RM-synthesis spectrum for each pixel). The grey scale is chosen to make structures with positive $\phi$ more visible. The contour line corresponds to $\phi=0$~\radm2. Four yellow lines, fitted by-eye, indicate the location of the four filaments that can be identified in this map, and which are located in the eastern part of the system of HI shells.}
  \label{FaradayRotationMap}
\end{figure}

\begin{figure}[tbp]
  \includegraphics[clip, width=10cm, angle=0, bb=55 305 625 713]{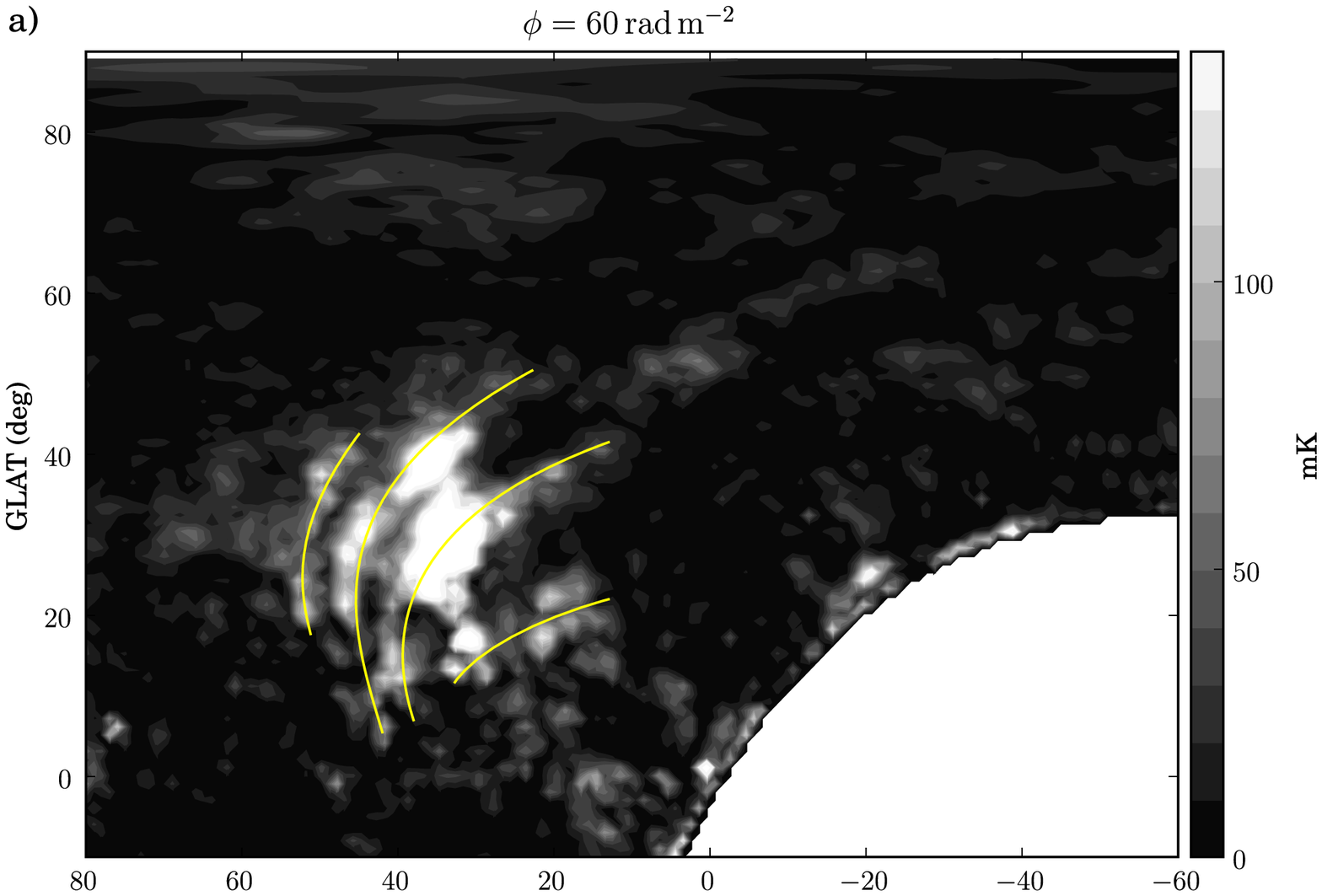}\\
  \includegraphics[clip, width=10cm, angle=0, bb=55 305 625 713]{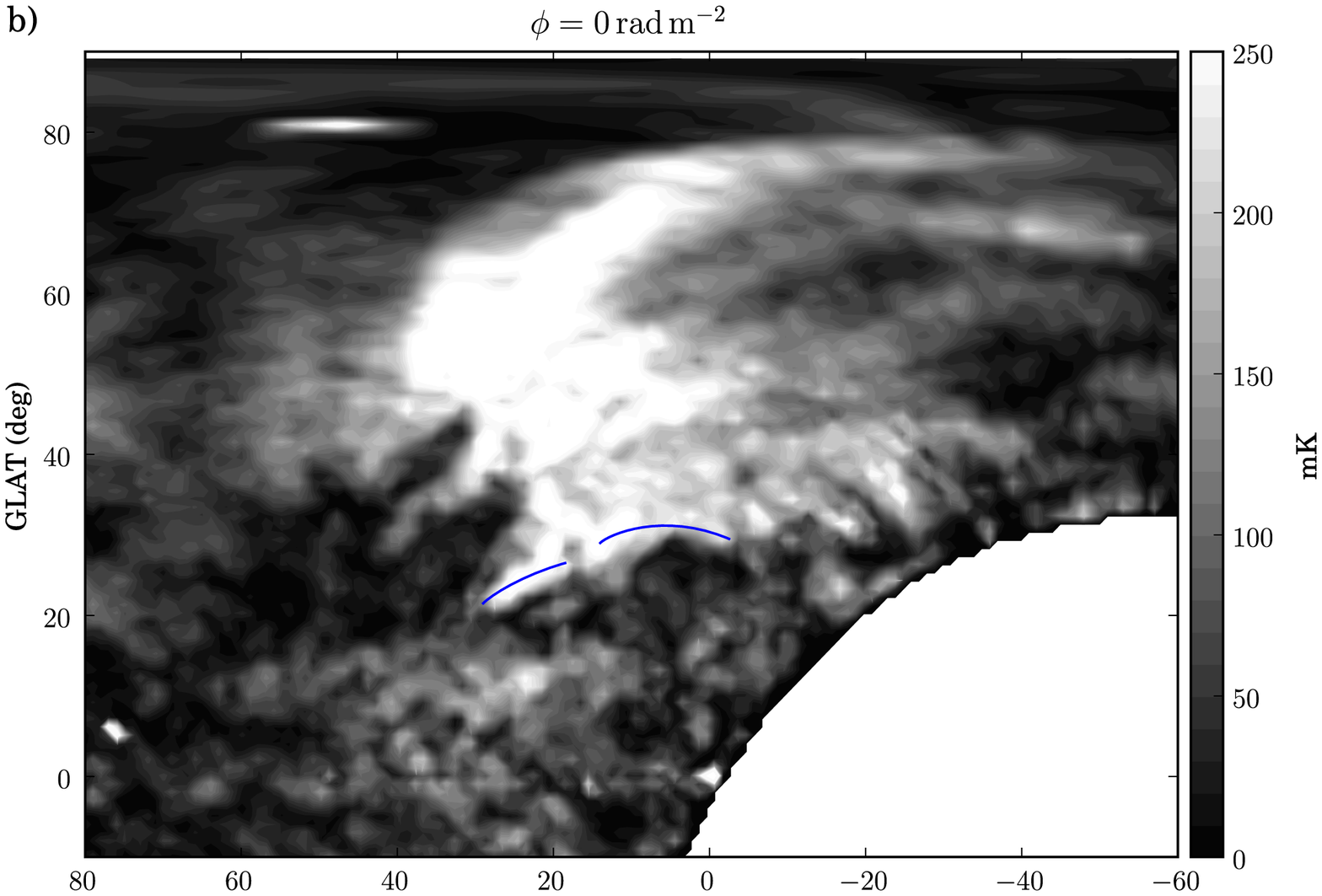}\\
  \includegraphics[clip, width=10cm, angle=0, bb=55 305 625 713]{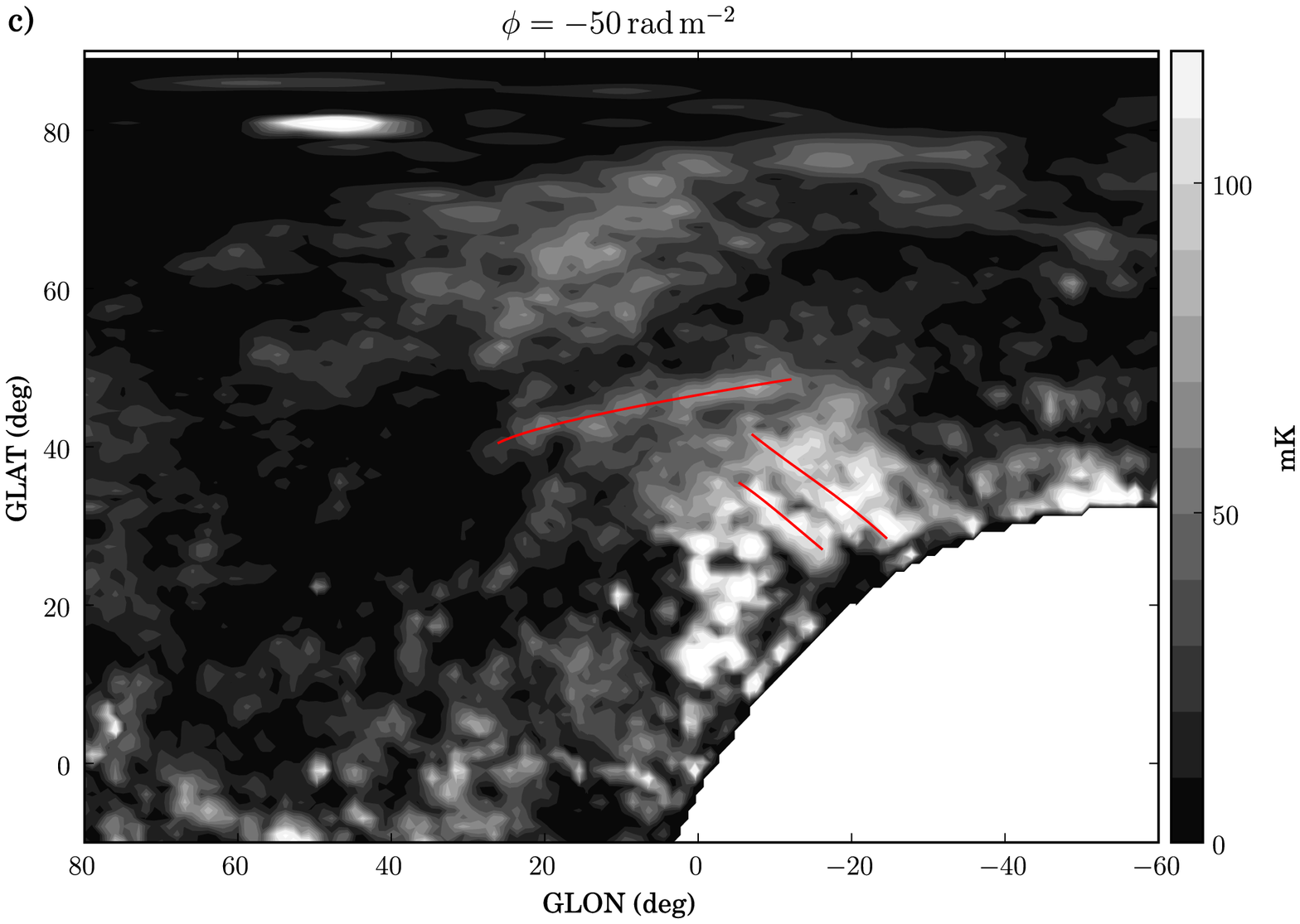}\\
  \caption{RM-synthesis frames showing polarized intensity at Faraday depths of 60~\radm2 (a), 0~\radm2 (b), and -50~\radm2 (c). The four yellow lines from Fig.~\ref{FaradayRotationMap} are repeated in (a). Blue and red lines in (b) and (c) indicate polarized filaments that were identified in RM-synthesis frames.}
  \label{RMSynthesisFrames}
\end{figure}

\begin{figure}[tbp]
  \includegraphics[clip, width=\columnwidth, angle=0, bb=55 307 667 712]{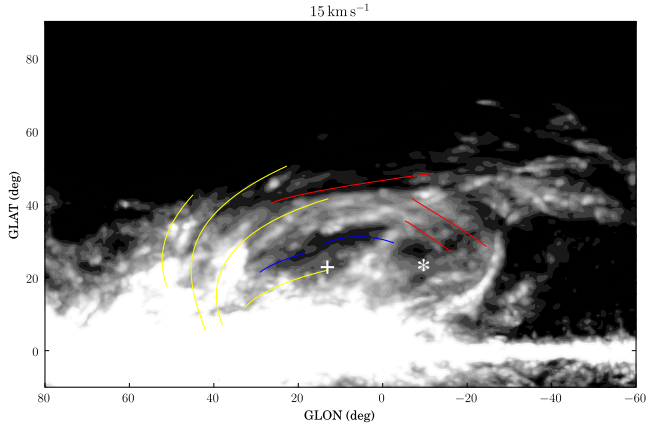}
	\label{fig:HI}
  \caption{Map of the HI brightness temperature at $v_\mathrm{lsr}=15$~$\kms$ in a logarithmic grey scale. The position of the polarized filaments, as well as the position of the Upper Sco subgroup (Sco OB2\_2) today (``\textasteriskcentered'') and 5 Myr ago (``+'') are shown. The brightness temperatures of the HI shells are about 3~K stronger than the background.}
  \label{HIMap}
\end{figure}

The polarization data presented in this paper are the first results from the Global Magneto-Ionic Medium Survey (GMIMS\footnote{\texttt{https://www.astrosci.ca/users/drao/gmims}}). The frequency coverage is 1277 to 1740 MHz with the band split into 2048 frequency channels. The angular resolution at the present sky coverage is $\sim 1^\circ$. The rms noise at this stage of the survey is 25 mK in a single channel, and is 1 mK in an RM-synthesis frame. The reader is referred to \citet{Wolleben:2009, Wolleben:2010} for more information on observing and calibration.

RM-synthesis was performed on data cubes of Stokes~$U$ and $Q$. Incomplete frequency coverage causes sidelobes in the Rotation Measure Spread Function (RMSF, the response function with which the RM-synthesis spectra are convolved) which were partially removed by applying a CLEAN algorithm to the data cube. The expected width of the RMSF is 132~\radm2 (note that structures at different Faraday depths can be distinguished with higher accuracy). By using a narrower RMSF for the cleaning process a final resolution of 60~\radm2 was achieved\footnote{We carefully checked that all conclusions derived from the super-resolved RM cube also apply to the un-CLEANED RM cube.}. 

Intrinsic polarization angles cannot be recovered accurately from our RM-Synthesis data due to the low S/N ratio of the emission associated with the bubble. For a detailed discussion see \citet{2005A&A...441.1217B}.

\subsection{Errors and uncertainties}

Systematic errors in the data are caused by two uncertainties in the calibration. First, the aperture efficiency of the DRAO 26-m Telescope, which is used to calculate main beam brightness temperatures in $U$ and $Q$, is known only at 1.4 GHz. For the data calibration we assumed this value to be constant over the band, neglecting possible variation of up to a few percent. The effect of a 20\% gradient of aperture efficiency over the band on the final RM-synthesis frames was tested and, while the systematic error in the polarized intensities was found to be $\lesssim 10\%$, the shape and position of polarized structures were not affected on a level relevant for this analysis. The second uncertainty is the absolute calibration (zero-levels) of $U$ and $Q$, which was derived by subtracting ground emission profiles obtained by averaging over a region of the sky relatively free of Galactic emission -- a method shown to work reasonably well for long scans, if confusion with Galactic emission can be avoided. 

In order to estimate the reliability of the calibrated data we compared polarized intensity (PI) with an older, precisely calibrated survey at 1.4 GHz \citep{2006A&A...448..411W}. We find a good correlation down to $\mbox{PI}\approx 100$~mK. Since the correlation can be expected to suffer from bandwidth depolarization in the wide-band data, we use this value as a conservative estimate of the reliability threshold in our RM-synthesis data. In the present analysis we consider only polarized structures above that threshold.

\subsection{Filaments of strong polarized emission associated with HI shells}

Figure~\ref{FaradayRotationMap} shows a Faraday rotation map obtained from the RM-synthesis cube. Shown in this map is $\phi$ of the dominant emission in each pixel (i.e. the Faraday depth of the strongest polarized emission, which is not necessarily the highest or lowest $\phi$ along that l.o.s.). The global properties of the Faraday rotation shown in this map resemble the large-scale pattern of pulsar and EGS RMs seen in the same field \citep[e.g.][]{1997A&A...322...98H, 2009ApJ...702.1230T} in both sign and magnitude of the Faraday rotation. The high sensitivity to local polarized emission as well as the unprecedented angular resolution of our data, however, allow us to also clearly identify small-scale structures. We see filamentary structures of high positive $\phi$ in the region $10^\circ<l<60^\circ$ and $0^\circ<b<+50^\circ$, embedded in a large region of positive RMs.

Figure~\ref{RMSynthesisFrames} shows three RM-synthesis maps at $\phi=+60$~\radm2, $0$~\radm2, and $-50$~\radm2. The emission in these maps was Faraday rotated by the corresponding $\phi$ somewhere along the l.o.s. Note that these are not the only RM-synthesis frames that contain polarized emission from this region: the frames shown in Fig.~\ref{RMSynthesisFrames} have been selected in order to highlight filamentary structures at different Faraday depths. However, there is polarized emission with Faraday depths up to $\pm 100\FRM$.

The filaments seen in Fig.~\ref{FaradayRotationMap} appear again in the $+60$~\radm2 map as emission structures but also in the 0~\radm2 map as depressions: the $\phi = 0$~\radm2 frame shows a ``negative imprint'' of the filaments and the amplitudes of the polarized emission in the two frames in this region is similar. The missing emission in the $0$~\radm2 frame has simply been shifted into the $+60$~\radm2 frame by the action of a Faraday screen. 

More polarized emission filaments are identified at $\phi=0$~\radm2 and $-50$~\radm2. The three
filaments in the $-50$~\radm2 map (red lines in Fig.~\ref{RMSynthesisFrames}c) are
strongest in polarized intensity at $\sim-20$~\radm2 but are most
easily distinguished from the background in the map shown here. 

Figure~\ref{HIMap} shows HI at a velocity of $15 \kms$, obtained from the LAB-HI survey \citep{2005A&A...440..775K} at a resolution of $\sim 0.6\degr$ for the same field as Fig.~\ref{FaradayRotationMap} and ~\ref{RMSynthesisFrames}. At this velocity, shells of HI emission forming a bubble (or bubbles) are clearly visible. Projected onto the sky the bubble is elongated in Galactic longitude with a major axis of about $70^\circ$, minor axis of about $40^\circ$, and centred at $l=10^\circ$ and $b=+25^\circ$. The filaments of polarized emission (Fig.~\ref{FaradayRotationMap} and ~\ref{RMSynthesisFrames}) closely follow the shape of the HI shells, with the connection becoming even clearer when scrolling through the HI emission in neighbouring velocity channels. 

In the following Sections we discuss the physical properties of this $2200\sqdeg$ HI bubble and deduce its magnetic field structure from our RM-synthesis cube. Note that the region we are discussing covers $1/20$ of the sky. 

\section{Distance to and Origin of the Filaments}

Large, extra-planar, expanding HI shells in the direction of the Galactic centre are associated with stellar winds and supernova remnants (SNR) originating in the stars of the Scorpius-Centaurus OB association \citep{1992A&A...262..258D}. Of the three identified subgroups in Sco-Cen, two have been identified with the origin of the Local Bubble and the Loop~I superbubble \citep{2001ApJ...560L..83M}. The third sub-group, Upper Scorpius (also know as Sco OB2\_2), is the furthest away at a distance of $145\pc$ \citep{1999AJ....117..354D} and its position of $l=-8\deg$, $b=+21\deg$ places it in the centre of the western side of the HI shells shown in Fig.~\ref{HIMap}. 5 Myr ago, when this subgroup formed, it was located at $l=15\degr$, $b=+21\degr$ \citep{2001ApJ...560L..83M}: the first OB stellar winds and SNR in the Upper Sco  subgroup, expected to occur when it was $3$--$5$ Myr old, will have occurred as it travelled along the major axis of the system of HI shells. Furthermore, 5 Myr ago the $14$--$15$ Myr old subgroup Upper Centaurus Lupus (UCL) was located at $l=4\degr$, $b=+9\degr$ about $100\pc$ away \citep{2001ApJ...560L..83M}. The HI filaments in the north-east of Fig.~\ref{HIMap} could plausibly have been swept up by SNR and winds from the UCL subgroup.

If the centre of the HI bubble is at a distance of $145\pc$ then its linear size is about $200\pc$ $\times$ $100\pc$. If the cross section along the l.o.s. is circular then the nearside of the bubble is $95\pc$ away and the far side $195\pc$ distant. These dimensions are compatible with other, independent, information that constrains the bubble size: since it shows no sign of breaking out of the disk its height above the mid-plane must be less than the warm HI scaleheight of about $300\pc$; the depolarizing effect of the $170\pc$ distant HII-region Sh2-27 (centre at $l=6.1\degr$, $b=+23.6\degr$, diameter of $11\degr$) can be seen in Fig~\ref{RMSynthesisFrames}(b) so most of the polarized emission in this region must originate beyond 170~pc; the wall of the local cavity is about $80\pc$ distant in this direction so the HI shells of this bubble should be no nearer than this.

Mixed emitting and Faraday-rotating regions produce polarized emission from a continuous range of Faraday depths because emission from the near edge undergoes little Faraday rotation whereas emission from the far side is rotated by the whole layer. The observed RM-spectra for l.o.s. in this region generally have a strong peak (with polarized intensities $>100$~mK), which we interpret as the Faraday rotation of the background polarized emission, but many l.o.s. show weak emission from a range of Faraday depth, indicating that low level ($\lesssim 50$~mK) emission might also be coming from the shell. The shell acts as a Faraday screen to the strong background polarized emission (thin in $\phi$) and also as a much weaker mixed emitting and rotating slab. No radio continuum emission from this bubble is seen in total intensity at 408 MHz \citep{1982A&AS...47....1H} and 1.4 GHz \citep{1982A&AS...48..219R, 1986A&AS...63..205R}, but $\sim 50$~mK emission is too low to stand out from the synchrotron background in these maps.

The Faraday depths of the emission rotated by the bubble allow us to determine its magnetic field configuration. Note that with $1\degr$ resolution all structures within the bubble that are bigger than about $3\pc$ are resolved. The positive $\phi$ in the east (Fig.~\ref{RMSynthesisFrames}a) changes to negative $\phi$ in the west (Fig.~\ref{RMSynthesisFrames}c), suggesting that the magnetic field is wrapped around the bubble, pointing towards us in the east and away from us in the west. The polarized emission at $\phi=0$ along the centre of the bubble links these two regions: $\phi=0$ means that either $\Brpa=0$ or $\n=0$. In this central region the magnetic field pattern indicated by optical polarization measurements of local stars \citep[][]{2000AJ....119..923H}, as well as radio polarization data at 1.4 GHz \citep[][]{2006A&A...448..411W}, show that $\Brpe$ closely follows the shape of the HI bubble. This suggests that, in this part of the shell, $\Brpa$ must be small with a $\Brpe$ component of unknown strength. The bottom of the bubble lies near the Galactic plane and is not clearly visible in any $\phi$ frame.

\begin{figure}[tbp]
  \includegraphics[clip, width=\columnwidth, angle=0, bb=15 18 272 243]{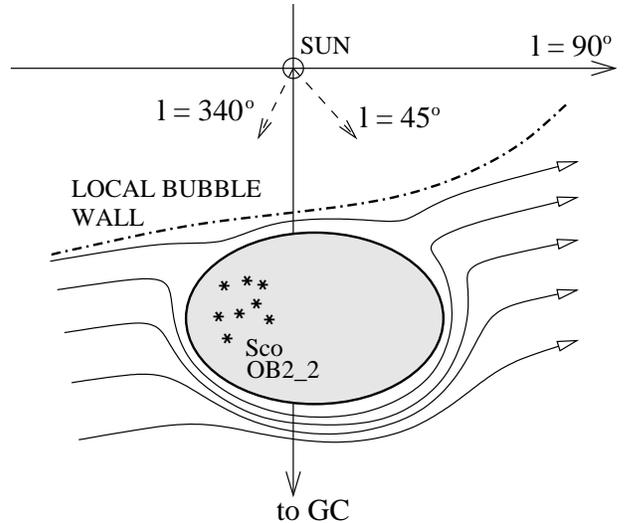}
  \caption{The sketch depicts the position and geometry of the HI shell and its magnetic field relative to the Local Bubble. Sco OB2\_2 is drawn at its current position. GC denotes Galactic centre.}
  \label{Sketch}
\end{figure}

Theoretical models of magnetic fields in interstellar bubbles symmetrically inflated by stellar winds and SNR \citep[e.g.][]{1991ApJ...375..239F, 1992PASJ...44..177T} predict that the ambient B-field is compressed in a shell by the expanding shock wave. The shape and implied magnetic field configuration of this HI bubble suggest that it has expanded asymmetrically. In particular, for $\Brpa$ to have constant magnitude but opposite signs at the two ends of the bubble indicates that its expansion is constrained to only one direction along the l.o.s. The dense wall of the Local Bubble lies in the right direction and at the right distance to constrain the expansion of shells powered by the Upper Scorpius sub-group (5-6 Myr old) towards us. It seems likely that magnetic pressure in the wall of the Local Bubble has caused an assymmetric expansion. Figure 4 shows a sketch of the deduced bubble and magnetic field configuration.

\section{Discussion}

We have searched the combined VTSS/SHASSA/WHAM survey \citep{2003ApJS..146..407F} for signs of the HI shells and bubble in H$\alpha$ emission but could not find any. Assuming no optical extinction, this suggests that the high Faraday rotation from the shells is caused by a local enhancement of magnetic field and not electron density. The emission measure towards the shells is around $2.4$~cm$^{-6}\,$pc. Assuming a scale height for the Warm Ionized Medium of 0.9~kpc \citep{Savage:2009} to 1.8~kpc \citep{Gaensler:2008} we have an average electron density along $b=20\degr$ of $0.03$--$0.02\cmcube$. The shells have an approximate thickness of $5\degr$ or $\delta r\approx 13\pc$, giving a maximum path length through a shell of $L\approx 2\sqrt{2r\delta r}\approx100\pc$. At the ends of the major axis of the bubble, where we expect the magnetic field to be mainly directed along the l.o.s., the observed Faraday depth of $|\phi|\approx 50$--$60\FRM$ corresponds to a magnetic field strength of $\Brpa\approx 20$--$34\uG$. It is noteworthy that the Faraday rotation pattern has been fractured into confined filaments; naively one would expect a Faraday rotation distribution which changes smoothly from positive on one side to negative on the other.

Such strong magnetic fields are difficult to reconcile with a simple sweeping up of a typical ambient $2$--$3\uG$ field by a single strong adiabatic shock, such as a SNR in the Sedov-Taylor phase, which has a compression ratio of only four. Two possibilities to explain the strong magnetic fields are: (i) the SNR and/or stellar winds that inflated the bubble have entered the radiative stage of their evolution (plausible given the 5--6 Myr age of the sub-group), allowing stronger compression to occur; (ii) the Upper Sco sub-group OB stars had already entered the dense wall of the local bubble, with its already compressed magnetic field of about 8-9$\uG$ \citep{2006ApJ...640L..51A, 2010ApJ...714.1170M}, when the new bubble expanded. The latter possibility is compatible with the elongated shape (Section 3).

The main astrophysical consequence of this nearby bubble and its associated Faraday rotation is its effect on the interpretation of EGS and pulsar Faraday rotation in terms of the global magnetic field structure of the Milky Way. 

Maps of pulsar RMs \citep[e.g.][]{1997A&A...322...98H} show a clear antisymmetry in the sign of RM in the northern Galactic hemisphere looking toward the Galactic centre, with positive RMs dominating in the region $60^\circ>l>0^\circ$, $b>0\degr$ and negative RMs at $0^\circ>l>-60^\circ$, $b>0$. The recent Faraday rotation map of NVSS sources by \citet{2009ApJ...702.1230T}, which provides the most densely sampled grid of RMs to date, shows a similar antisymmetric pattern of RMs. The RMs of pulsars and EGS in this region have magnitudes up to $100\FRM$, similar to the maximum Faraday depths found in our RM-synthesis cube.

It is tempting to interpret such large-scale antisymmetries in Faraday rotation as a signature of the regular magnetic field of the Milky Way \citep{1988Ap.....28..247A, 1997A&A...322...98H, 2008A&A...477..573S}, especially as galactic dynamo theory makes predictions about the expected mean field symmetries in the disk and halo \citep[e.g.][]{Beck:1996}. But several authors have pointed out that the antisymmetry may be due to radio Loop~I and its bright, polarized segment, the NPS. \citet{1980ApJ...242...74S} find that the effect of the NPS on the RMs in the northern Galactic hemisphere is clearly visible and \citet{1981PASJ...33..603I} point out the importance of avoiding the effect of irregularities in RMs due to local objects. 

In this paper we have identified a nearby bubble and set of shells that covers a large area of the sky with the same Faraday rotation asymmetry as seen in the EGS and pulsar RMs. The discovery of small-scale structures in the Faraday rotation sky which are associated with nearby HI, have the same sign, similar magnitude, and are embedded in the large-scale antisymmetry at $b>0^\circ$ leads us to conclude that the observed antisymmetry has largely a local origin. Note that the structures discussed here are not associated with the NPS as they do not resemble the shape of the spur: thus the NPS can only add to the region of the northern sky where local structures dominate the Faraday rotation pattern. To minimize the effect of local RM structures on the picture of the mean Galactic magnetic field we suggest that point source RMs towards regions that show diffuse polarized emission at high Faraday depth should not be used.

Clearly the diffuse polarized radio emission contains important information about the \textit{local} magneto-ionic medium, with better angular resolution than currently available from EGS source measurements. This must be taken into account when building a complete picture of the Galactic magnetic field. At present RM-synthesis data exist only for the northern sky. Observations for a southern sister-survey (STAPS, principal investigator: M. Haverkorn) were just completed and RM-Synthesis should soon become available too. \citet{1992A&A...262..258D} (in their Fig. 1b) show more HI filaments associated with Sco OB2\_2 in the southern sky. Our prediction is that there will be Faraday rotation structures also associated with these filaments.

\acknowledgements{}

We thank JinLin Han for useful comments on the manuscript, Anvar Shukurov for discussions, and Fred Gent for an instructive movie based on his numerical simulations. We used the online tools VizieR and SkyView. AF is grateful for the STFC grant ST/F003080/1. B.M.G acknowledges the support of the Australian Research Council through grant FF0561298. The Dominion Radio Astrophysical Observatory is a National Facility operated by the National Research Council Canada.

\end{document}